\documentclass[11pt]{article}

\usepackage{arxiv}

\usepackage[utf8]{inputenc}
\usepackage[T1]{fontenc}
\usepackage{url}
\usepackage{booktabs}
\usepackage{amsfonts}
\usepackage{nicefrac}
\usepackage{microtype}
\usepackage{lipsum}
\usepackage{graphicx}
\graphicspath{ {./images/} }
\usepackage{amsmath}
\usepackage{amssymb}
\usepackage{amsthm}
\usepackage[round,authoryear]{natbib}

\usepackage{longtable}
\usepackage{array}
\usepackage{multirow}
\usepackage{wrapfig}
\usepackage{float}
\usepackage{threeparttable}
\usepackage{threeparttablex}
\usepackage[normalem]{ulem}
\usepackage{xcolor}

\newtheorem{lemma}{Lemma}

\newcommand{\LM}{{\normalfont \mathrm{LM}}}\newcommand{\OLS}{{\normalfont \mathrm{OLS}}}
\newcommand{\LMM}{{\normalfont \mathrm{LMM}}}

\title{Improving External Prediction in New Clusters with Mixed-Effects Models: Implications for Multicenter Studies}

\author{
  Jiawa Zhang \\
  Department of Biostatistics, Mailman School of Public Health \\
  Columbia University, New York, NY 10027, U.S.A. \\
  \And
  Hyungrok Do \\
  Division of Biostatistics, Department of Population Health \\
  NYU Grossman School of Medicine, New York, NY 10016, U.S.A. \\
  \And
  Myeonggyun Lee \\
  Division of Biostatistics, Department of Population Health \\
  NYU Grossman School of Medicine, New York, NY 10016, U.S.A. \\
  \texttt{myeonggyun.lee@nyulangone.org}
}

\begin{document}
\maketitle

\begin{abstract}
    Prediction models developed from multicenter data are often applied to new centers. A linear mixed model (LMM) captures between-center heterogeneity through random effects. However, a new center’s random effect is unknown, so the LMM predicts from fixed effects alone, as does a standard linear model (LM) that ignores clustering. The two models thus differ only in how fixed effects are estimated, and it is unclear whether the LMM predicts better. We decompose the prediction error in a new center into fixed-effect estimation error, between-center variation, and residual variation. Because the last two components are identical for both models, any LMM advantage must come from more precise fixed-effect estimates. In simulations across diverse designs and random-effect structures, the LMM generally achieved lower mean squared error than the LM. Relative gains increased with the number of predictors, between-center heterogeneity, and imbalance in center sizes, and decreased as training data grew. Modeling a random slope yielded additional gains that increased with between-center variation in the predictor effect. In a stress test using held-out eICU hospitals selected for strong outcome heterogeneity, a random-intercept LMM predicted ICU length of stay more accurately than an LM. These findings clarify the conditions under which mixed models can improve prediction in new centers and can inform model development for multicenter studies.
    
\end{abstract}

\keywords{clustered data \and external prediction \and generalized least squares \and linear mixed models \and prediction error}

\section{Introduction}

Clinical prediction models are often developed from data collected in several hospitals or studies and then used in settings that did not contribute to model development \citep{wynants2019untapped}. Such models often perform worse in new settings than in the data used to develop them \citep{siontis2015external}, and external validation in new centers is therefore recommended before a model is used \citep{steyerberg2016prediction, riley2016external}. Data pooled from several centers are clustered by center, yet a review of 50 multicenter cardiovascular risk prediction studies found that 39 ignored this clustering in the statistical analysis \citep{wynants2019untapped}.

For a continuous outcome, ignoring clustering typically means fitting a standard linear model (LM) to the pooled data. For example, the CKD-EPI equation for estimating glomerular filtration rate, a routine measure of kidney function, was developed by linear regression on data pooled from ten studies \citep{levey2009new}. A linear mixed-effects model (LMM) accounts for clustering by adding center-specific random effects to the LM, so that the outcome level and the effects of predictors can vary between centers \citep{laird1982random, verbeke2000linear}. Mixed-effects models are standard tools for clustered data \citep{raudenbush2002hierarchical, goldstein2011multilevel}, and they can also be used for prediction \citep{skrondal2009prediction, bouwmeester2013clustered}. Which of the two models predicts better, however, depends on whether the patient comes from a center that contributed to model development or from a new center.

The key difference is whether outcome data from the patient's center are available \citep{skrondal2009prediction}. If the center contributed to model development, its observed outcomes are available, and the LMM can use them to predict the center's random effect. The usual choice is the best linear unbiased predictor (BLUP), and adding it to the fixed-effects part gives a center-specific prediction \citep{henderson1975best, robinson1991blup, jianglahiri2006mixed}. A new center, in contrast, did not contribute to model development, so none of its outcomes are available for predicting its random effect. Without other information about the new center, the LMM sets the center's random effect to its mean of zero and therefore predicts from the fixed effects alone, as the LM does. The two models then differ only in how they estimate the fixed effects. This leads to our research question: when a model for a continuous outcome is applied to new centers, does the LMM’s estimation of the fixed effects yield more accurate predictions than the LM’s, and under what conditions?

Existing work does not answer this question directly. Several methods improve on the fixed-effects prediction by using information about the new center when it is available \citep{finkelman2016dynamic, ni2018informative, jiang2018classified}. Other work concerns how the prediction model is built. It has examined how to choose intercepts or predictors that transfer well across studies \citep{debray2013framework, dejong2021generalizable}, extended the mixed model with tree-based mean structures \citep{gottard2023trees}, and studied how a misspecified random-effects distribution affects prediction \citep{vu2026misspecification}. These approaches either need information about the new center, which is often unavailable when a model is first used there, or change how the model is specified. They do not address the choice between an LM and an LMM itself.

The studies closest to our question compared standard and mixed-effects logistic models for binary outcomes, and they distinguish predictions that use the estimated cluster effects from predictions that do not. \citet{bouwmeester2013clustered} found that a random-intercept logistic model discriminated better than a standard logistic model only when its predictions included the estimated cluster effects. In new clusters, where these effects are unknown, the two models discriminated similarly. Calibration also depends on whether the predictions include center effects and on whether it is assessed within centers or in the population as a whole \citep{wynants2018ignoring}. A later simulation study found that random-intercept logistic models were better calibrated than standard logistic models, including in new centers \citep{falconieri2020risk}. Together, these studies show that accounting for clustering affects predictive performance. For a continuous outcome, however, performance is usually measured by the mean squared prediction error (MSPE), and these studies do not show whether an LMM reduces this error in a new center.

Statistical theory gives only a partial answer. The LM estimates the fixed effects by ordinary least squares (OLS), whereas the LMM uses generalized least squares (GLS), which accounts for the correlation among patients in the same center. When the variance components are known, GLS is the best linear unbiased estimator \citep{aitken1936least} and is therefore at least as precise as OLS. In practice, the variance components are unknown and are usually estimated by maximum likelihood or restricted maximum likelihood \citep{harville1977maximum}. Estimating them adds variability to the GLS estimates, especially in small samples \citep{kackar1984approximations, kenward1997small}. It is therefore unclear how much of the advantage of GLS remains in typical multicenter data, and whether it translates into smaller prediction errors in new centers.

In this article, we show that an LMM can improve prediction in a new center even though the new center's random effect is unknown. The reason is clear from a decomposition of the MSPE in a new center into fixed-effect estimation error, between-center variation, and residual variation. Only the first term differs between the two models, so the LMM can help only by estimating the fixed effects more precisely. Simulations with random intercepts and random slopes quantify this benefit, with all models evaluated in clusters not used for fitting \citep{royston2004construction, roberts2017crossvalidation}. An analysis of ICU length of stay in the eICU Collaborative Research Database \citep{pollard2018eicu} provides a stress test, in which the hospitals with the most atypical mean length of stay are held out. The LMM generally predicted better than the LM, with the largest gains when the fixed effects were hard to estimate. Section~\ref{sec:methods} presents the decomposition, Sections~\ref{sec:simulation} and~\ref{sec:eicu} report the simulations and the eICU application, respectively, and Section~\ref{sec:discussion} discusses practical implications and limitations.

\section{Methods}\label{sec:methods}

We consider the standard linear mixed-effects model (LMM) \citep{searle1992variance,verbeke2000linear}
\begin{equation}
    y_{ij} = \mathbf{x}_{ij}^{T} \boldsymbol{\beta} + \mathbf{z}_{ij}^{T} \mathbf{b}_i + \varepsilon_{ij},
    \qquad i = 1, \ldots, n, \quad  j = 1, \ldots, m_i,
    \label{eq:lmm}
\end{equation}
where $y_{ij}$ denotes the continuous outcome for observation $j$ in cluster $i$, $\mathbf{x}_{ij} \in \mathbb{R}^p$ is the vector of fixed-effect covariates, and $\boldsymbol{\beta} \in \mathbb{R}^p$ represents the corresponding population-level regression coefficients. The term $\mathbf{z}_{ij} \in \mathbb{R}^q$ is a vector of random-effect covariates associated with cluster-specific deviations $\mathbf{b}_i$, assumed to follow a mean-zero multivariate normal distribution $\mathbf{b}_i \sim \mathcal{N}(\mathbf{0}, D)$, where $D \in \mathbb{R}^{q \times q}$ is a positive semidefinite covariance matrix. The within-cluster residual errors $\varepsilon_{ij}$ are independently distributed as $\mathcal{N}(0, \sigma^2)$ and are assumed independent of the random effects $\mathbf{b}_i$.

For cluster $i$, let $X_i \in \mathbb{R}^{m_i \times p}$ and $Z_i \in \mathbb{R}^{m_i \times q}$ denote the fixed- and random-effects design matrices obtained by stacking $\mathbf{x}_{ij}^{T}$ and $\mathbf{z}_{ij}^{T}$ over $j=1,\ldots,m_i$. Let
\[
X = \begin{pmatrix} X_1 \\ \vdots \\ X_n \end{pmatrix} \in \mathbb{R}^{N \times p},
\qquad
Z = \operatorname{blockdiag}(Z_1,\ldots,Z_n) \in \mathbb{R}^{N \times nq},
\]
where $N = \sum_{i=1}^{n}m_i$. Stacking the cluster-specific random effects as $\mathbf{b}=(\mathbf{b}_1^T,\ldots,\mathbf{b}_n^T)^T$ gives $\operatorname{Var}(\mathbf{b})=I_n\otimes D$. Therefore, the marginal covariance matrix of the full response vector $\mathbf{y}=(y_{11},\ldots,y_{nm_n})^T$ is
\begin{equation}
V = Z(I_n \otimes D)Z^T + \sigma^2 I_N
  = \operatorname{blockdiag}_{i=1,\ldots,n}\left(Z_i D Z_i^T + \sigma^2 I_{m_i}\right),
\end{equation}
where $I_N$ is the $N \times N$ identity matrix. Under the correctly specified model \eqref{eq:lmm}, the generalized least squares (GLS) estimator of the fixed effects is
\begin{equation}
    \widehat{\boldsymbol{\beta}}_{\mathrm{LMM}} = (X^{T} V^{-1} X)^{-1} X^{T} V^{-1} \mathbf{y},
    \label{eq:lmm_est}
\end{equation}
which coincides with the maximum likelihood (or restricted maximum likelihood) estimator under Gaussian assumptions. For comparison, the ordinary least squares (OLS) estimator that ignores random effects, equivalently, fitting a standard linear model (LM) to the pooled data, is
\begin{equation}
    \widehat{\boldsymbol{\beta}}_{\mathrm{OLS}} = (X^{T} X)^{-1} X^{T} \mathbf{y}.
    \label{eq:ols_est}
\end{equation}

The two estimators differ only in their weighting of observations through the covariance structure $V$. The LMM estimator~\eqref{eq:lmm_est} appropriately accounts for within-cluster correlation and between-cluster heterogeneity via $V^{-1}$, whereas the OLS estimator~\eqref{eq:ols_est} used by the LM gives all observations their usual least-squares weighting. This difference leads to a corresponding difference in MSPE.
\begin{equation}
\Sigma_{\widehat{\boldsymbol{\beta}},\mathrm{LMM}} = (X^{T} V^{-1} X)^{-1},
\qquad\Sigma_{\widehat{\boldsymbol{\beta}},\mathrm{OLS}} = (X^{T} X)^{-1} X^{T} V X (X^{T} X)^{-1}.
\end{equation}

Let an external cluster have size $m'$ with design matrices $X' \in \mathbb{R}^{m' \times p}$ and $Z' \in \mathbb{R}^{m' \times q}$. Consider the response vector for this cluster:\begin{equation}\mathbf{y}' = X'\boldsymbol{\beta} + Z'\mathbf{b}' + \boldsymbol{\varepsilon}',\quad \mathbf{b}' \sim \mathcal{N}(\mathbf{0}, D),\quad \boldsymbol{\varepsilon}' \sim \mathcal{N}(\mathbf{0}, \sigma^{2} I_{m'}),\end{equation}where $\mathbf{b}'$ and $\boldsymbol{\varepsilon}'$ are independent of the training data used to estimate $\widehat{\boldsymbol{\beta}}$.

\begin{lemma}[Cluster-Level MSPE]\label{lem:lmm-vs-lm-cluster-formal}
Assume the fixed-effects mean is correctly specified.\par
For predictions in the external cluster, let
\[
\widehat{y}' = X'\widehat{\boldsymbol{\beta}}.
\]
The mean squared prediction error satisfies:
\begin{equation}
\begin{aligned}
    \mathrm{MSPE}_{\LMM} &= \frac{1}{m'}\mathbb{E}\big[ \lVert \mathbf{y}' - X'\widehat{\boldsymbol{\beta}}_{\LMM} \rVert_{2}^{2} \big] \\
    &= \frac{1}{m'}\operatorname{tr}\left( X' \Sigma_{\widehat{\boldsymbol{\beta}},\mathrm{LMM}} (X')^{T} \right)
     + \frac{1}{m'}\operatorname{tr}\left( Z' D (Z')^{T} \right)
    + \sigma^{2}, \\
    \mathrm{MSPE}_{\LM} &= \frac{1}{m'}\mathbb{E}\big[ \lVert \mathbf{y}' - X'\widehat{\boldsymbol{\beta}}_{\OLS} \rVert_{2}^{2} \big] \\
    &= \frac{1}{m'}\operatorname{tr}\left( X' \Sigma_{\widehat{\boldsymbol{\beta}},\mathrm{OLS}} (X')^{T} \right)
    + \frac{1}{m'}\operatorname{tr}\left( Z' D (Z')^{T} \right)
    + \sigma^{2}.
\end{aligned}
\end{equation}
By the Gauss–Markov theorem, $\Sigma_{\widehat{\boldsymbol{\beta}},\mathrm{LMM}} \preceq \Sigma_{\widehat{\boldsymbol{\beta}},\mathrm{OLS}}$ in the Loewner order. Consequently, for every $X'$:
\begin{equation}
    \operatorname{tr}\left( X' \Sigma_{\widehat{\boldsymbol{\beta}},\mathrm{LMM}} (X')^{T} \right)\leq\operatorname{tr}\left( X' \Sigma_{\widehat{\boldsymbol{\beta}},\mathrm{OLS}} (X')^{T} \right).
\end{equation}
Equality holds if $D = 0$ or if the training design satisfies the conditions for OLS efficiency (e.g., specific balanced designs). 
\end{lemma}

\begin{proof}
The prediction error decomposes as
\begin{equation}
    \mathbf{y}' - X'\widehat{\boldsymbol{\beta}}= X'(\boldsymbol{\beta} - \widehat{\boldsymbol{\beta}})+ Z'\mathbf{b}'+ \boldsymbol{\varepsilon}'.
\end{equation}
Taking the squared Euclidean norm and expectation, we utilize the independence of the training estimator $\widehat{\boldsymbol{\beta}}$ from the new cluster's random effects $\mathbf{b}'$ and errors $\boldsymbol{\varepsilon}'$. Since all terms have zero means, the cross-terms vanish:
\begin{equation}
\begin{aligned}
    \mathbb{E}\big[ \lVert \mathbf{y}' - X'\widehat{\boldsymbol{\beta}} \rVert_{2}^{2} \big]
    &= \mathbb{E}\big[ \lVert X'(\boldsymbol{\beta} - \widehat{\boldsymbol{\beta}}) \rVert_{2}^{2} \big]
    + \mathbb{E}\big[ \lVert Z'\mathbf{b}' \rVert_{2}^{2} \big]
    + \mathbb{E}\big[ \lVert \boldsymbol{\varepsilon}' \rVert_{2}^{2} \big].
\end{aligned}
\end{equation}
Dividing this identity by $m'$ and using the trace trick $\mathbb{E}[\mathbf{u}^T \mathbf{u}] = \operatorname{tr}(\mathbb{E}[\mathbf{u}\mathbf{u}^T])$, the first term's contribution to MSPE becomes:
\begin{equation}
    \frac{1}{m'}\mathbb{E}\big[ (\boldsymbol{\beta} - \widehat{\boldsymbol{\beta}})^T (X')^T X' (\boldsymbol{\beta} - \widehat{\boldsymbol{\beta}}) \big]= \frac{1}{m'}\operatorname{tr}\left( X' \operatorname{Var}(\widehat{\boldsymbol{\beta}}) (X')^{T} \right).
\end{equation}
Similarly, the second term contributes $m'^{-1}\operatorname{tr}(Z' D (Z')^T)$ and the third contributes $\sigma^2$. Substituting the variances derived from the training data:
\begin{equation}
    \operatorname{Var}(\widehat{\boldsymbol{\beta}}_{\LMM}) = \Sigma_{\widehat{\boldsymbol{\beta}},\LMM},\qquad\operatorname{Var}(\widehat{\boldsymbol{\beta}}_{\OLS}) = \Sigma_{\widehat{\boldsymbol{\beta}},\OLS}.
\end{equation}
Since $\widehat{\boldsymbol{\beta}}_{\LMM}$ is the Best Linear Unbiased Estimator (BLUE), we have $\Sigma_{\widehat{\boldsymbol{\beta}},\mathrm{LMM}} \preceq \Sigma_{\widehat{\boldsymbol{\beta}},\mathrm{OLS}}$. Multiplying the Loewner inequality on both sides by $X'$ and $(X')^{T}$ preserves the ordering, and taking the trace yields the claimed inequality for the prediction errors.
\end{proof}

\section{Simulation Study}\label{sec:simulation}

The simulation study examined whether LMMs improve prediction in new clusters compared with LMs under finite sample conditions, and when this improvement is most evident. Simulation I included a random intercept and compared an LM with an LMM containing a random intercept. Simulation II included both a random intercept and one random slope. It compared an LM, an LMM containing only a random intercept, and a correctly specified LMM containing both random effects. This design separated the benefit of accounting for baseline cluster heterogeneity from the additional benefit of allowing a covariate effect to vary across clusters. We varied the training sample size, number of clusters, number of predictors, balance in cluster sizes, and strength of random effect heterogeneity. Performance was evaluated in entirely new clusters using predictions based only on the estimated fixed effects. The primary measure was the mean squared error (MSE) of the predictions, an empirical estimate of the MSPE. MSE was calculated within each test cluster and then averaged with equal weight across clusters. We also reported the percentage reduction in MSE achieved by each LMM relative to the LM.

\subsection{Simulation Design}

Training data contained $N$ observations from $K$ clusters and $P$ covariates, excluding the intercept. Both simulations used
\begin{equation}
Y_{ij}=\beta_0+\mathbf{x}_{ij}^{T}\boldsymbol\beta+b_{0i}+b_{1i}X_{ij1}+\varepsilon_{ij},
\end{equation}
with
\begin{equation}
b_{0i}\sim N(0,\tau_0^2),\qquad \varepsilon_{ij}\sim N(0,1),\qquad
\begin{cases}
b_{1i}=0, & \text{Simulation I},\\
b_{1i}\sim N(0,\tau_1^2),\ \operatorname{Cov}(b_{0i},b_{1i})=0, & \text{Simulation II}.
\end{cases}
\end{equation}
Each covariate was generated independently from a standard normal distribution. We set $\beta_0=1$, $\beta_1=\beta_2=1$, and $\beta_r=0$ for $r\geq3$, with residual standard deviation $\sigma=1$.

Simulation I fitted LM and LMM-I. Simulation II fitted LM, LMM-I, and LMM-IS. Hereafter, LMM-I denotes the random-intercept model and LMM-IS denotes the random-intercept-and-random-slope model. LMM-I modeled only baseline variation across clusters, whereas LMM-IS also allowed the coefficient of the first covariate to vary across clusters. Although the simulated intercept and slope were independent, their covariance was estimated rather than fixed during model fitting. Mixed models were fitted by maximum likelihood.

Under the balanced design, each cluster contained $m=N/K$ observations. Under structural imbalance, cluster sizes increased in proportion to the cluster index $i$. Counts were rounded with a minimum of two per cluster, and any discrepancy was corrected among the largest clusters so that the total equaled $N$. Because allocation depended only on cluster index, cluster size was independent of covariates, random effects, and outcomes.

For the training data, we varied the total sample size $N$ in two ways. We first added clusters while keeping 20 observations per cluster, and then added observations within a fixed set of $K=10$ clusters. When another factor was varied, the reference training setting was $N=200$, $K=10$, $P=10$, and $\tau_0=2$, with $\tau_1=1$ in Simulation II. Table~\ref{tab:design} separates the training design, external test design, and simulation implementation.

\begin{table}[htbp]
\centering
\caption{Simulation settings for the training data, external test data, and Monte Carlo implementation.}
\label{tab:design}
\small
\renewcommand{\arraystretch}{1.08}
\begin{tabular}{@{}>{\raggedright\arraybackslash}p{0.27\textwidth}>{\raggedright\arraybackslash}p{0.09\textwidth}>{\raggedright\arraybackslash}p{0.25\textwidth}>{\raggedright\arraybackslash}p{0.31\textwidth}@{}}
\toprule
Setting & Symbol & Values examined & Design details \\
\midrule
\multicolumn{4}{@{}l}{\textit{Training data}} \\
Increase $N$ by adding clusters & $N$ & 100, 200, 500, 1000 & $m=20$; $K=N/20$ \\
Increase $N$ within clusters & $N$ & 50, 200, 500, 1000 & $K=10$; $m=N/10$ \\
Vary number of clusters & $K$ & 5, 10, 20, 40 & Total sample size fixed at $N=200$ \\
Vary number of covariates & $P$ & 3, 10, 20, 50 & Intercept not included in $P$ \\
Vary random-intercept SD & $\tau_0$ & 0.5, 1, 1.5, 2 & Simulations I and II \\
Vary random-slope SD & $\tau_1$ & 0.5, 1, 1.5, 2 & Simulation II only \\
Residual SD & $\sigma$ & 1 & Fixed in all scenarios \\
Cluster-size pattern & $m_j$ & Balanced or structurally unbalanced & Unbalanced sizes proportional to $1,\ldots,K$ \\
Reference training setting & --- & $N=200$, $K=10$, $P=10$, $\tau_0=2$ & $\tau_1=1$ in Simulation II; used when another factor was varied \\
\addlinespace
\multicolumn{4}{@{}l}{\textit{External test data}} \\
Number of new clusters & $K_{\mathrm{test}}$ & 50 & Generated independently of the training data \\
Observations per new cluster & $m_{\mathrm{test}}$ & 1000 & Equal across test clusters \\
\addlinespace
\multicolumn{4}{@{}l}{\textit{Simulation implementation}} \\
Monte Carlo replicates & $S$ & 1000 & Complete replicates for each scenario \\
\bottomrule
\end{tabular}
\end{table}

External prediction was evaluated in an independently generated test sample containing 50 new clusters with 1000 observations per cluster. Predictions from all models used fixed effects only. For replicate $s$ and model $a$, the primary metric was MSE,
\begin{equation}
\operatorname{MSE}_{s}(a)
=
\frac{1}{K_{\mathrm{test}}}
\sum_{i=1}^{K_{\mathrm{test}}}
\left\{
\frac{1}{m_{\mathrm{test}}}
\sum_{j=1}^{m_{\mathrm{test}}}
\left(
Y_{ij}^{\mathrm{test}}
-
\widehat{Y}_{ij,a}^{\mathrm{test}}
\right)^2
\right\}.
\end{equation}
This MSE gives equal weight to each unseen cluster. Because the test clusters had equal sizes, it was numerically equal to observation-weighted MSE but retained the interpretation of prediction for a typical unseen cluster.

For model $a\in\{\mathrm{LMM\mbox{-}I},\mathrm{LMM\mbox{-}IS}\}$, relative improvement from LM was defined as
\begin{equation}
\mathrm{RI}(a)=\frac{100}{S}\sum_{s=1}^{S}
\frac{\operatorname{MSE}_{s}(\mathrm{LM})-\operatorname{MSE}_{s}(a)}
{\operatorname{MSE}_{s}(\mathrm{LM})}.
\end{equation}
In Simulation I, we reported $\mathrm{RI}(\mathrm{LMM\mbox{-}I})$; in Simulation II, we reported both $\mathrm{RI}(\mathrm{LMM\mbox{-}I})$ and $\mathrm{RI}(\mathrm{LMM\mbox{-}IS})$. Their difference describes the incremental improvement from modeling the random slope beyond using a random intercept. Each scenario included 1000 complete replicates. Replicates with fitting errors or missing performance values were regenerated; singular fits that returned valid predictions were retained.

\subsection{Simulation Results}

\subsubsection{Random-intercept data-generating setting}

Across all scenarios generated from the random-intercept model, LMM-I had lower MSE than LM under both balanced and structurally unbalanced training designs (Table~\ref{tab:ri-results} and Web Figure~\ref{fig:web-ri}). This consistent ordering supports the proposed mechanism: accounting for covariance within the training clusters improves estimation of the population fixed effects, even though prediction in a new cluster cannot use a cluster-specific random intercept.

Prediction error decreased as the training information increased. With $K=10$, the largest small-sample gain occurred at $N=50$, where LMM-I reduced MSE by 12.47\% under balance and 13.02\% under imbalance. The gain became smaller as $N$ increased. At the same total sample size, distributing observations across more clusters generally produced a lower LMM-I MSE than adding observations within 10 existing clusters. Increasing $K$ from 5 to 40 at $N=200$ also reduced LMM-I MSE from 5.90 to 5.21. These patterns indicate that information from additional independent clusters is particularly valuable for estimating population effects.

The separation between LM and LMM-I increased with the number of predictors and the strength of cluster heterogeneity. As $P$ increased from 3 to 50, RI increased from 1.06\% to 16.66\% under balance and from 2.51\% to 16.92\% under imbalance. Increasing $\tau_0$ raised the MSE of both models because variability in a new cluster is irreducible, but it also increased the relative gain from LMM-I. Thus, larger $P$ and $\tau_0$ did not necessarily lower absolute prediction error; instead, they made modeling the training covariance more important.

The scenario that maximized performance depended on the criterion. The lowest LMM-I MSE occurred at $\tau_0=0.5$, with values of 1.33 under balance and 1.34 under imbalance, because this setting contained the least between-cluster variation. The largest improvement relative to LM occurred at $P=50$, where RI reached 16.66\% and 16.92\%, respectively. LMM-I therefore achieved its lowest absolute error in the easiest prediction setting but its greatest comparative advantage in the highest-dimensional setting.

Structural imbalance strengthened the advantage of LMM-I. In every paired scenario in Table~\ref{tab:ri-results}, RI was larger under imbalance than under balance, and Web Figure~\ref{fig:web-ri} shows a wider separation between the corresponding LM and LMM-I curves. Unequal cluster sizes concentrate the training information in a smaller number of clusters. Accounting for this covariance structure therefore provides a greater benefit than treating all observations as independent.

\begin{table}[htbp]
\centering
\caption{MSE for LM and LMM-I when data were generated from the random-intercept model. For both balanced and structurally unbalanced training designs, MSE results are presented as mean (empirical SD), and relative improvement (RI) is presented as the mean percentage across 1000 replicates.}
\label{tab:ri-results}
\scriptsize
\resizebox{\textwidth}{!}{%
\begin{tabular}{@{}llrrr rrr@{}}
\toprule
&&\multicolumn{3}{c}{Balanced training clusters}&\multicolumn{3}{c}{Structurally unbalanced training clusters}\\
\cmidrule(lr){3-5}\cmidrule(lr){6-8}
Scenario&Setting&LM&LMM-I&RI (\%)&LM&LMM-I&RI (\%)\\
\midrule
Training sample size ($N/K=20$)&$N=100$&6.24 (1.48)&5.87 (1.40)&5.81&6.45 (1.79)&5.88 (1.42)&7.54\\
&$N=200$&5.64 (1.02)&5.45 (1.02)&3.37&5.74 (1.14)&5.45 (1.02)&4.67\\
&$N=500$&5.26 (0.85)&5.19 (0.85)&1.55&5.31 (0.87)&5.19 (0.85)&2.28\\
&$N=1000$&5.09 (0.79)&5.05 (0.79)&0.77&5.11 (0.81)&5.05 (0.79)&1.22\\
\addlinespace
Training sample size ($K=10$)&$N=50$&6.72 (1.40)&5.82 (1.11)&12.47&6.79 (1.51)&5.83 (1.11)&13.02\\
&$N=200$&5.66 (1.06)&5.47 (1.04)&3.41&5.77 (1.14)&5.47 (1.04)&4.86\\
&$N=500$&5.48 (0.99)&5.40 (0.99)&1.38&5.59 (1.14)&5.40 (0.98)&2.84\\
&$N=1000$&5.48 (1.12)&5.45 (1.12)&0.69&5.57 (1.22)&5.45 (1.12)&1.81\\
\addlinespace
Number of clusters ($N=200$)&$K=5$&6.07 (1.49)&5.90 (1.47)&2.96&6.27 (1.76)&5.90 (1.47)&4.70\\
&$K=10$&5.57 (1.05)&5.38 (1.03)&3.44&5.67 (1.15)&5.39 (1.04)&4.54\\
&$K=20$&5.48 (0.89)&5.28 (0.87)&3.71&5.52 (0.92)&5.28 (0.88)&4.28\\
&$K=40$&5.41 (0.87)&5.21 (0.86)&3.70&5.42 (0.87)&5.21 (0.86)&3.87\\
\addlinespace
Number of covariates&$P=3$&5.48 (1.05)&5.42 (1.05)&1.06&5.60 (1.20)&5.42 (1.05)&2.51\\
&$P=10$&5.66 (1.07)&5.46 (1.07)&3.53&5.74 (1.15)&5.46 (1.06)&4.54\\
&$P=20$&5.95 (1.05)&5.55 (1.01)&6.76&6.07 (1.17)&5.55 (1.01)&8.01\\
&$P=50$&6.97 (1.23)&5.79 (1.06)&16.66&7.03 (1.40)&5.79 (1.06)&16.92\\
\addlinespace
Random-intercept SD&$\tau_0=0.5$&1.34 (0.08)&1.33 (0.07)&0.73&1.35 (0.08)&1.34 (0.08)&0.91\\
&$\tau_0=1.0$&2.22 (0.27)&2.17 (0.26)&2.00&2.23 (0.30)&2.17 (0.27)&2.53\\
&$\tau_0=1.5$&3.63 (0.59)&3.53 (0.58)&2.95&3.70 (0.64)&3.53 (0.58)&4.18\\
&$\tau_0=2.0$&5.61 (1.07)&5.42 (1.05)&3.30&5.70 (1.16)&5.42 (1.05)&4.51\\
\bottomrule
\end{tabular}}
\end{table}
\subsubsection{Random-intercept and random-slope setting}

The ordering was consistent across the displayed scenarios: LMM-IS had the lowest MSE, LMM-I was intermediate, and LM had the highest error (Table~\ref{tab:rias-results} and Web Figure~\ref{fig:web-ris}). With $K=10$ and $N=50$, the balanced design produced improvements of 9.40\% for LMM-I and 11.81\% for LMM-IS relative to LM. Under structural imbalance, the corresponding improvements were 10.17\% and 12.83\%. Thus, the random intercept captured most of the gain in this small-sample scenario, while the random slope provided a further improvement.

The two RI columns clarify when the random slope added material predictive value. Under balance, increasing $\tau_1$ from 0.5 to 2 changed $\mathrm{RI}_{\mathrm{I}}$ from 3.23\% to 1.76\%, while $\mathrm{RI}_{\mathrm{IS}}$ increased from 3.46\% to 4.01\%. Under structural imbalance, the corresponding pairs changed from 4.24\% versus 4.52\% to 2.41\% versus 5.35\%. When slope heterogeneity was weak, the random-intercept model performed nearly as well as the correctly specified model. As slope heterogeneity increased, modeling the slope accounted for a larger share of the reducible error.

Total MSE increased with $\tau_1$ because slope variation in an unseen cluster is irreducible. Correctly modeling that variation in the training clusters nevertheless improved estimation of the common fixed effects. This distinction is practically important: a random intercept can provide a robust baseline when a high-dimensional model cannot support many random slopes, while carefully selected random slopes can improve prediction when there is strong evidence that specific effects vary across clusters.

\begin{table}[htbp]
\centering
\caption{MSE and reductions versus LM for LMM-I and LMM-IS under balanced and structurally unbalanced training designs. Values are mean (empirical SD) across 1000 replicates.}
\label{tab:rias-results}
\scriptsize
\resizebox{\textwidth}{!}{%
\begin{tabular}{@{}llrrrrr rrrrr@{}}
\toprule
&&\multicolumn{5}{c}{Balanced training clusters}&\multicolumn{5}{c}{Structurally unbalanced training clusters}\\
\cmidrule(lr){3-7}\cmidrule(lr){8-12}
&&\multicolumn{1}{c}{LM}&\multicolumn{2}{c}{LMM-I}&\multicolumn{2}{c}{LMM-IS}&\multicolumn{1}{c}{LM}&\multicolumn{2}{c}{LMM-I}&\multicolumn{2}{c}{LMM-IS}\\
\cmidrule(lr){3-3}\cmidrule(lr){4-5}\cmidrule(lr){6-7}\cmidrule(lr){8-8}\cmidrule(lr){9-10}\cmidrule(lr){11-12}
Scenario&Setting&MSE (SD)&MSE (SD)&\shortstack{Reduction\\vs LM (\%)}&MSE (SD)&\shortstack{Reduction\\vs LM (\%)}&MSE (SD)&MSE (SD)&\shortstack{Reduction\\vs LM (\%)}&MSE (SD)&\shortstack{Reduction\\vs LM (\%)}\\
\midrule
Training sample size ($N/K=20$)&$N=100$&7.55 (1.57)&7.19 (1.50)&4.64&7.09 (1.50)&5.93&7.84 (1.84)&7.34 (1.63)&5.61&7.20 (1.59)&7.27\\
&$N=200$&6.79 (1.11)&6.60 (1.10)&2.72&6.55 (1.09)&3.51&6.98 (1.24)&6.69 (1.10)&3.72&6.61 (1.08)&4.81\\
&$N=500$&6.34 (0.89)&6.26 (0.89)&1.23&6.24 (0.89)&1.61&6.40 (0.97)&6.28 (0.92)&1.75&6.25 (0.92)&2.29\\
&$N=1000$&6.19 (0.84)&6.15 (0.84)&0.63&6.14 (0.84)&0.83&6.17 (0.83)&6.11 (0.82)&0.87&6.10 (0.82)&1.16\\
\addlinespace
Training sample size ($K=10$)&$N=50$&8.05 (1.49)&7.23 (1.20)&9.40&7.03 (1.14)&11.81&8.10 (1.55)&7.20 (1.24)&10.17&6.98 (1.15)&12.83\\
&$N=200$&6.81 (1.12)&6.62 (1.10)&2.78&6.56 (1.10)&3.62&6.90 (1.19)&6.62 (1.07)&3.68&6.55 (1.05)&4.70\\
&$N=500$&6.66 (1.11)&6.59 (1.10)&1.15&6.57 (1.10)&1.48&6.75 (1.19)&6.58 (1.06)&2.16&6.54 (1.05)&2.80\\
&$N=1000$&6.55 (1.03)&6.52 (1.03)&0.56&6.50 (1.02)&0.75&6.71 (1.18)&6.57 (1.06)&1.70&6.53 (1.05)&2.26\\
\addlinespace
Number of clusters ($N=200$)&$K=5$&7.33 (1.68)&7.17 (1.66)&2.26&7.12 (1.65)&2.85&7.57 (1.86)&7.24 (1.60)&3.57&7.14 (1.58)&4.76\\
&$K=10$&6.85 (1.12)&6.65 (1.10)&2.89&6.60 (1.10)&3.64&6.89 (1.14)&6.61 (1.06)&3.74&6.53 (1.05)&4.93\\
&$K=20$&6.55 (0.90)&6.36 (0.90)&2.82&6.30 (0.89)&3.72&6.61 (1.00)&6.38 (0.95)&3.22&6.32 (0.94)&4.13\\
&$K=40$&6.46 (0.91)&6.28 (0.89)&2.79&6.22 (0.89)&3.67&6.48 (0.91)&6.29 (0.89)&2.97&6.22 (0.88)&3.94\\
\addlinespace
Number of covariates&$P=3$&6.61 (1.11)&6.55 (1.11)&0.81&6.53 (1.10)&1.15&6.73 (1.24)&6.57 (1.12)&2.02&6.52 (1.11)&2.60\\
&$P=10$&6.78 (1.05)&6.59 (1.03)&2.79&6.53 (1.02)&3.62&6.96 (1.17)&6.67 (1.09)&3.78&6.59 (1.08)&4.87\\
&$P=20$&7.18 (1.06)&6.76 (1.02)&5.69&6.65 (1.01)&7.27&7.31 (1.25)&6.82 (1.13)&6.29&6.69 (1.11)&8.03\\
&$P=50$&8.35 (1.27)&7.17 (1.10)&13.85&6.86 (1.08)&17.48&8.46 (1.47)&7.21 (1.12)&14.13&6.90 (1.11)&17.78\\
\addlinespace
Random-intercept SD&$\tau_0=0.5$&2.50 (0.30)&2.49 (0.30)&0.27&2.44 (0.28)&2.56&2.52 (0.31)&2.51 (0.31)&0.40&2.44 (0.28)&3.06\\
&$\tau_0=1.0$&3.34 (0.40)&3.30 (0.40)&1.24&3.24 (0.38)&2.87&3.40 (0.41)&3.33 (0.39)&1.70&3.25 (0.36)&3.96\\
&$\tau_0=1.5$&4.79 (0.67)&4.70 (0.66)&1.96&4.65 (0.65)&3.02&4.88 (0.74)&4.72 (0.68)&2.98&4.64 (0.65)&4.52\\
&$\tau_0=2.0$&6.78 (1.09)&6.59 (1.08)&2.81&6.53 (1.07)&3.66&6.92 (1.24)&6.64 (1.13)&3.64&6.56 (1.12)&4.78\\
\addlinespace
Random-slope SD&$\tau_1=0.5$&5.93 (1.06)&5.74 (1.05)&3.23&5.73 (1.05)&3.46&6.10 (1.20)&5.81 (1.06)&4.24&5.79 (1.06)&4.52\\
&$\tau_1=1.0$&6.89 (1.21)&6.70 (1.20)&2.70&6.64 (1.19)&3.59&6.90 (1.18)&6.65 (1.09)&3.35&6.57 (1.08)&4.51\\
&$\tau_1=1.5$&8.29 (1.26)&8.11 (1.23)&2.13&7.97 (1.20)&3.74&8.41 (1.37)&8.13 (1.26)&3.05&7.93 (1.20)&5.32\\
&$\tau_1=2.0$&10.28 (1.50)&10.09 (1.48)&1.76&9.86 (1.44)&4.01&10.40 (1.70)&10.13 (1.60)&2.41&9.80 (1.47)&5.35\\
\bottomrule
\end{tabular}}
\vspace{0.3em}
\parbox{0.98\textwidth}{\scriptsize Reductions compare each mixed-effects model with LM. Grouping MSE and reduction beneath each model is intended to make the model-wise comparisons easier to scan. $P$ excludes the intercept.}
\end{table}

\section{Application to eICU Hospital Data}\label{sec:eicu}

\subsection{Data source and analytic cohort}

We used the eICU Collaborative Research Database, a publicly available multicenter critical care database \citep{pollard2018eicu}. It contains data from hospitals with different patient populations, care practices, and institutional settings. This variation reflects a common challenge in practice. A model trained in several hospitals may later be used in a new hospital with little local outcome data. Reliable prediction of ICU length of stay in such hospitals may inform bed planning, staffing, and resource allocation \citep{murray2023forecasting}. We therefore used eICU to compare predictions from LM and a random-intercept LMM (LMM-I) in unseen hospitals. Specifically, we examined whether accounting for between-hospital heterogeneity during training improves external prediction.

The analytic file contained 53,561 ICU stays from 206 hospitals and 65 original columns. Hospital identifier defined the cluster, and ICU length of stay in hours was the outcome. We excluded 33 identifier or administrative fields, timestamps, post-discharge variables, duplicate outcomes, derived mortality predictions, variables with more than 25\% missingness, and variables unsuitable for a parsimonious model, including a diagnosis field with more than 500 levels.

The final predictor set contained 25 continuous and four categorical variables. Indicator coding produced 37 fixed-effect terms, excluding the intercept. The outcome was the natural logarithm of ICU length of stay. After defining the hospital sets, all preprocessing quantities were estimated from the training hospitals. The deidentified age category above 89 years was coded as 90. Missing continuous values were replaced with training-hospital medians, missing categorical values were coded as \emph{Unknown}, and categorical levels occurring in less than 3\% of the training data were combined as \emph{Other}. Continuous predictors were standardized using training-hospital means and standard deviations. These same quantities were applied to the external hospitals, preventing information leakage.

To compare LM and LMM-I under substantial hospital heterogeneity, we ranked hospitals by the absolute deviation of their mean log length of stay from the overall mean:
\begin{equation}
D_h=\left|\bar y_h-\bar y\right|,\qquad
\bar y_h=\frac{1}{n_h}\sum_{i=1}^{n_h}y_{hi}.
\end{equation}
The 20 most deviating hospitals formed the external test set, and the next 10 formed the training pool. This outcome-informed construction was designed to represent the conditions under which the simulations predicted the clearest LMM-I advantage. It is not an estimate of average generalizability across all eICU hospitals. Characteristics of the resulting hospital sets are summarized in Table~\ref{tab:eicu-cohort}.

\begin{table}[htbp]
\centering
\caption{Characteristics of the full cohort and hospital sets used for model development and external evaluation.}
\label{tab:eicu-cohort}
\small
\begin{tabular}{@{}p{0.27\textwidth}>{\centering\arraybackslash}p{0.18\textwidth}>{\centering\arraybackslash}p{0.27\textwidth}>{\centering\arraybackslash}p{0.20\textwidth}@{}}
\toprule
Characteristic & Full cohort & Training pool & External test set\\
\midrule
Role & Source population & Development and internal validation & External evaluation only\\
Rank by hospital deviation & Not applicable & 21--30 & 1--20\\
Hospitals & 206 & 10 & 20\\
ICU stays & 53,561 & 813 & 626\\
Hospital size, median (range) & 137.5 (1--1,983) & 56.0 (2--298) & 18.5 (2--90)\\
ICC for log length of stay & 0.0287 & 0.2170 & 0.5689\\
Mean ICU length of stay, hours & 128.3 & 120.4 & 155.4\\
\bottomrule
\end{tabular}
\end{table}

\subsection{Models and validation}

LM and LMM-I used the same outcome and 37 fixed-effect terms. LM was fitted by ordinary least squares. LMM-I added a hospital-specific random intercept and was fitted by maximum likelihood. We used one stratified 10-fold cross-validation of the 813 training-pool stays with a fixed random seed. Within each hospital, stays were randomly ordered and assigned in turn to folds. Each fitting set therefore retained observations from every training hospital, allowing its random intercept to be estimated for internal prediction.

For internal evaluation, we used held-out stays from hospitals represented in the corresponding fitting data. LM predictions used the estimated regression coefficients, whereas LMM-I predictions combined the fixed effects with the hospital-specific random intercept estimated from the other nine folds. For prediction error $r_{hi}^{(k)}=y_{hi}-\widehat y_{hi}^{(k)}$, MSE was first calculated within each hospital and then averaged with equal weight across hospitals, consistent with the simulation study.

For external evaluation, each fold-specific fitted model was applied to all 626 stays in the 20 hospitals excluded from model fitting. LMM-I predictions used fixed effects only because no random intercept could be estimated for an unseen hospital. External MSE weighted the 20 hospitals equally. We calculated MSE in every fold, summarized its mean and standard deviation, reported the percentage reduction from LM to LMM-I, and counted folds in which LMM-I had lower MSE.

\subsection{Results}

\begin{figure}[htbp]
\centering
\includegraphics[width=0.92\textwidth]{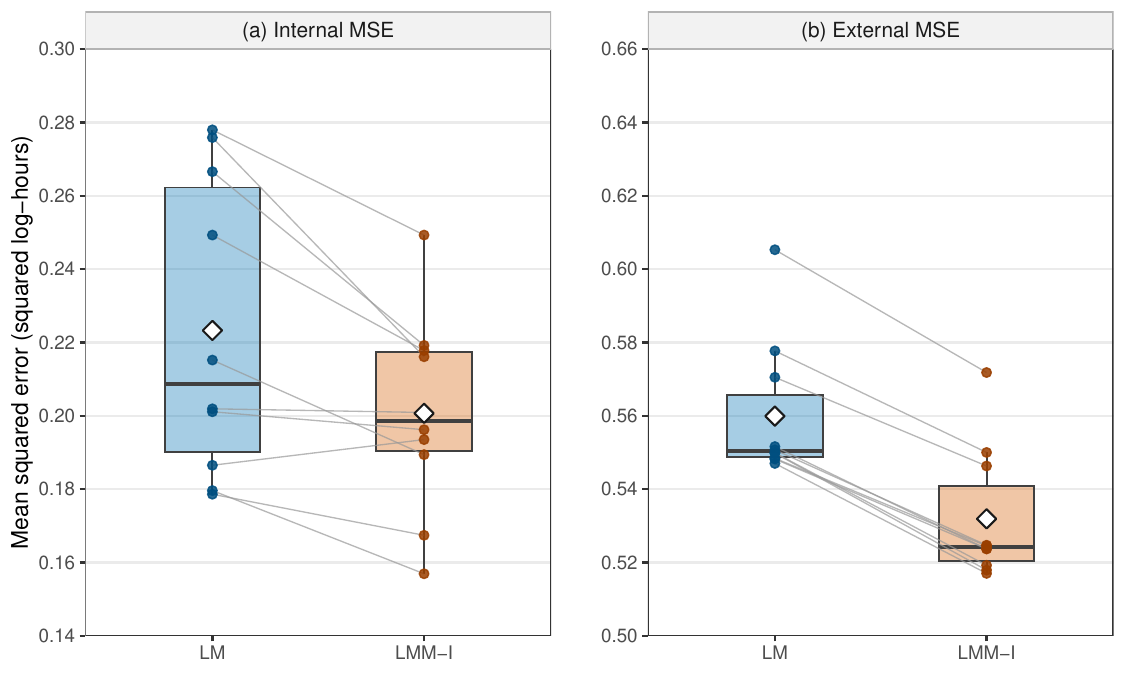}
\caption{Fold-level MSE from stratified 10-fold cross-validation. Gray lines connect LM and LMM-I values from the same fold; diamonds denote means across folds. Panels use a common 0.16 y-axis span and 0.02 increments, with different origins. Values are mean (SD) hospital-weighted MSE across 10 folds, expressed in squared log-hours. External (marginal) predictions: LM, 0.5599 (0.0191); LMM-I, 0.5319 (0.0181); relative improvement, 5.00\%; LMM-I lower in 10/10 folds. Internal (conditional) predictions: LM, 0.2233 (0.0403); LMM-I, 0.2007 (0.0268); relative improvement, 10.13\%; LMM-I lower in 9/10 folds. MSE was computed within hospital and averaged equally across hospitals. Conditional LMM-I predictions include estimated hospital random intercepts; marginal predictions use fixed effects only.}
\par\noindent\textbf{Alt text:} Paired LM and random-intercept LMM (LMM-I) fold-level MSE for internal and external hospitals. Gray lines link the same fold, and diamonds mark means. Both panels use 0.02 increments; ranges are 0.14--0.30 internally and 0.50--0.66 externally. Mean MSE is lower for LMM-I in both panels.
\label{fig:eicu}
\end{figure}

For external hospitals, marginal LMM-I predictions reduced the hospital-weighted MSE from 0.5599 (SD 0.0191) to 0.5319 (SD 0.0181). This corresponds to an absolute reduction of 0.0280 and a 5.00\% relative improvement. LMM-I had lower MSE in all 10 folds (Figure~\ref{fig:eicu}). The SD changed only slightly, from 0.0191 to 0.0181. Thus, the main benefit was a consistent reduction in average prediction error, not a large reduction in fold-to-fold variation. All folds showed improvement even though no local outcome data were available for the external hospitals. This pattern supports the proposed mechanism: modeling covariance during development can improve estimation of a population relationship that transfers to new hospitals.\par For internal hospitals, conditional LMM-I predictions reduced MSE from 0.2233 (SD 0.0403) to 0.2007 (SD 0.0268). The absolute reduction was 0.0226, and the relative improvement was 10.13\%. LMM-I had lower MSE in 9 of 10 folds. The SD also decreased substantially, from 0.0403 to 0.0268. This suggests that LMM-I reduced average error and produced more stable results across folds. For this task, the model could estimate a hospital-specific random intercept from the other nine folds.\par The absolute reductions were similar for the internal and external analyses. However, the relative improvement was about twice as large internally because the internal LM baseline MSE was lower. It does not indicate an absolute gain twice as large. The two analyses also represent different prediction tasks. Internal predictions used hospital-specific information, whereas external predictions used fixed effects only. Their MSE values therefore should not be interpreted as estimates of the same performance quantity.

Overall, these results show value in accounting for hospital-level clustering during model development. This approach may improve predictions when a model is transferred to a new hospital. It may also improve prediction stability within hospitals included in model development. More reliable length-of-stay predictions could help with bed-capacity, staffing, and resource planning. This may be especially useful at hospitals with little local outcome data. However, MSE was measured in squared log-hours. The observed gains therefore cannot be translated directly into changes in bed-days, staffing needs, or patient outcomes. The external hospitals were also selected to emphasize heterogeneity, and the training pool was small. Thus, the 5.00\% external gain provides evidence from a stress test. It should not be treated as the improvement expected at a typical hospital.

\section{Discussion}\label{sec:discussion}
Prediction in a completely new cluster differs from prediction within clusters represented during model development. Because no local outcome data are available, a new cluster's random effect cannot be estimated. Both LM and LMM predictions therefore rely on population-level fixed effects. The relevant question is thus not whether an LMM can predict the new cluster-specific deviation, but whether modeling clustered data during development improves estimation of the population relationship used for external prediction.

Our theoretical results addressed this question by decomposing external MSPE into fixed-effect estimation error, new-cluster random-effect variation, and residual variation. The latter two components are shared by LM and LMM when predictions use fixed effects only. Under the correctly specified mixed model, any LMM advantage arises from more efficient estimation of the fixed effects through covariance-aware weighting of the training data. This distinction is important because prior work has emphasized cluster-specific prediction, adaptation using local data, or cluster-level validation, whereas less is known about the direct benefit of modeling covariance for prediction in entirely unobserved clusters without target-cluster information.

The simulations were consistent with this mechanism. Under the data-generating settings considered, LMMs generally had lower external MSE than LMs, with larger relative gains when training data were limited, cluster heterogeneity was substantial, cluster sizes were unbalanced, or the fixed-effect model included more predictors. These gains should not be interpreted as recovery of a new cluster's random effect: between-cluster variation remains irreducible unless local outcome data become available. Similarly, the additional benefit of random slopes was most evident when slope heterogeneity was substantial. A random-intercept model may therefore be a useful starting point, while random slopes should be supported by scientific rationale, adequate numbers of clusters, and model diagnostics.

In the eICU analysis, the random-intercept LMM reduced hospital-weighted external MSE by 5.0\% relative to LM, despite using fixed effects only in held-out hospitals. This result is consistent with the proposed mechanism but should be interpreted cautiously. The external hospitals were selected to represent substantial outcome heterogeneity, the training pool was small, and cross-validation folds were not independent external replications. Thus, this analysis illustrates a setting in which covariance-aware estimation may help rather than estimating the average benefit expected across all hospitals.

Several limitations should be noted. The theoretical comparison assumes correct mean and covariance specification and treats variance components as known, whereas these quantities must be estimated in practice. In addition, random effects were assumed independent of covariates; when this assumption is implausible, within--between decompositions or correlated-random-effects approaches may be preferable \citep{mundlak1978pooling}. Our results also concern continuous outcomes under an identity-link model. Extensions to generalized mixed models, time-to-event outcomes, informative cluster size, covariance misspecification, and covariate shift across sites warrant further study.

In conclusion, for deployment in new clusters without local outcome data, an LMM cannot provide cluster-specific adaptation, but it can improve external prediction by estimating population-level effects more efficiently when the clustered covariance structure is appropriately modeled. External performance should therefore be evaluated by holding out entire clusters, and study design should prioritize the inclusion of diverse clusters rather than relying only on larger samples within existing clusters.

\section*{Acknowledgments}
During the preparation of this work, the authors used ChatGPT (OpenAI) for language editing and assistance with manuscript formatting. The authors reviewed and edited the output and take full responsibility for the content of the publication.

\section*{Funding}
The authors received no specific funding for this work.

\section*{Conflict of Interest}
The authors declare no conflicts of interest.

\section*{Data Availability Statement}
The eICU Collaborative Research Database analyzed in this study is available through PhysioNet to credentialed users who complete the required training and agree to the data use agreement.

\bibliographystyle{plainnat}
\bibliography{references}

\clearpage

\section*{Supplementary Material}

\begin{center}
\setcounter{figure}{0}\includegraphics[width=0.92\textwidth]{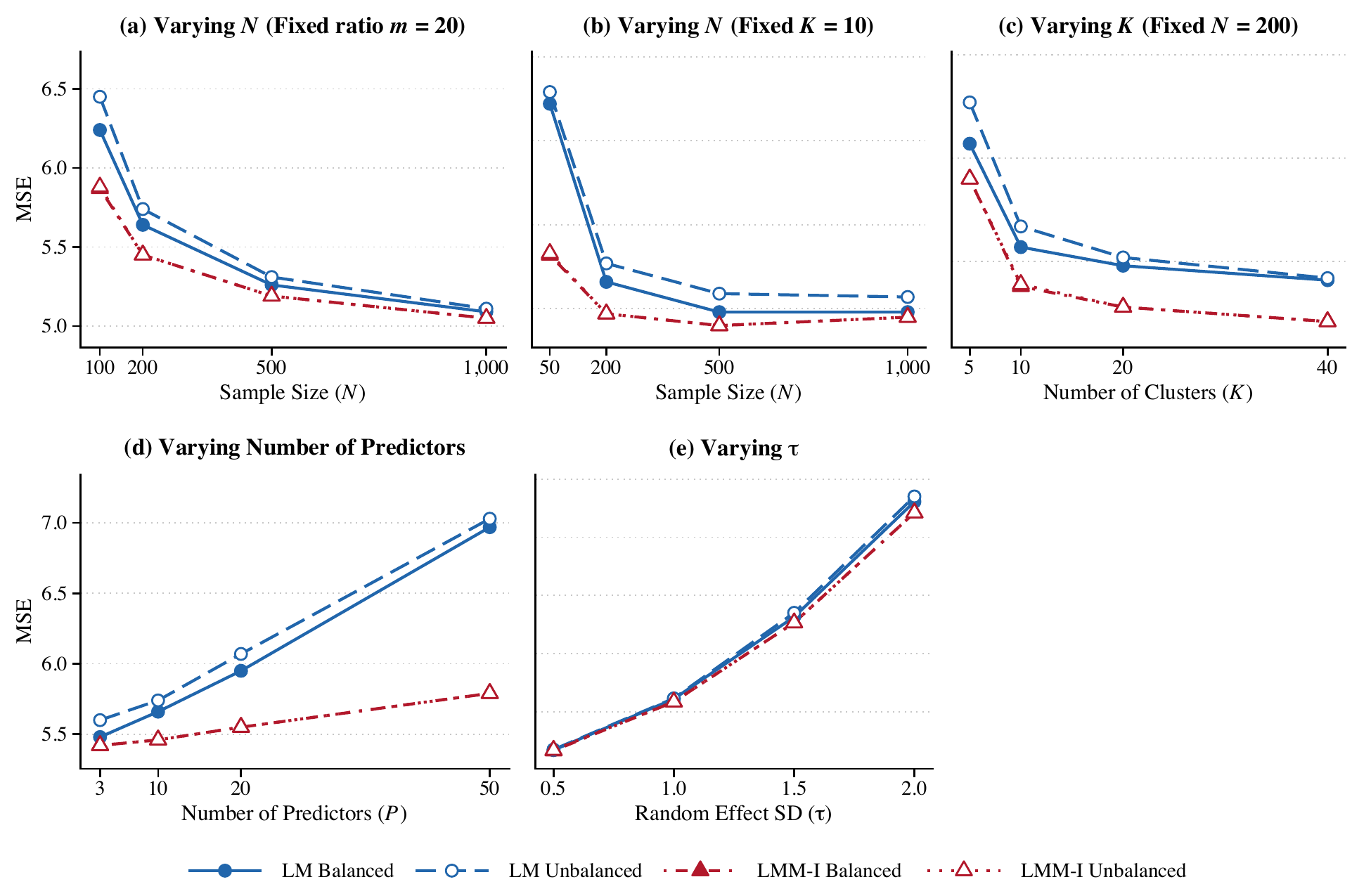}
\end{center}
\refstepcounter{figure}\label{fig:web-ri}\noindent\textbf{Web Figure \thefigure.} MSE for LM and LMM-I when data were generated from the random-intercept model. Points show Monte Carlo means, and error bars show 95\% Monte Carlo confidence intervals.

\par\noindent\textbf{Alt text:} Multi-panel comparison of LM and random-intercept LMM prediction MSE across training sample sizes, numbers of clusters and covariates, random-intercept variability, and balanced versus unbalanced cluster sizes. The LMM has lower MSE throughout, with the largest separation for smaller samples, more covariates, and unbalanced clusters.

\clearpage
\begin{center}
\includegraphics[width=0.92\textwidth]{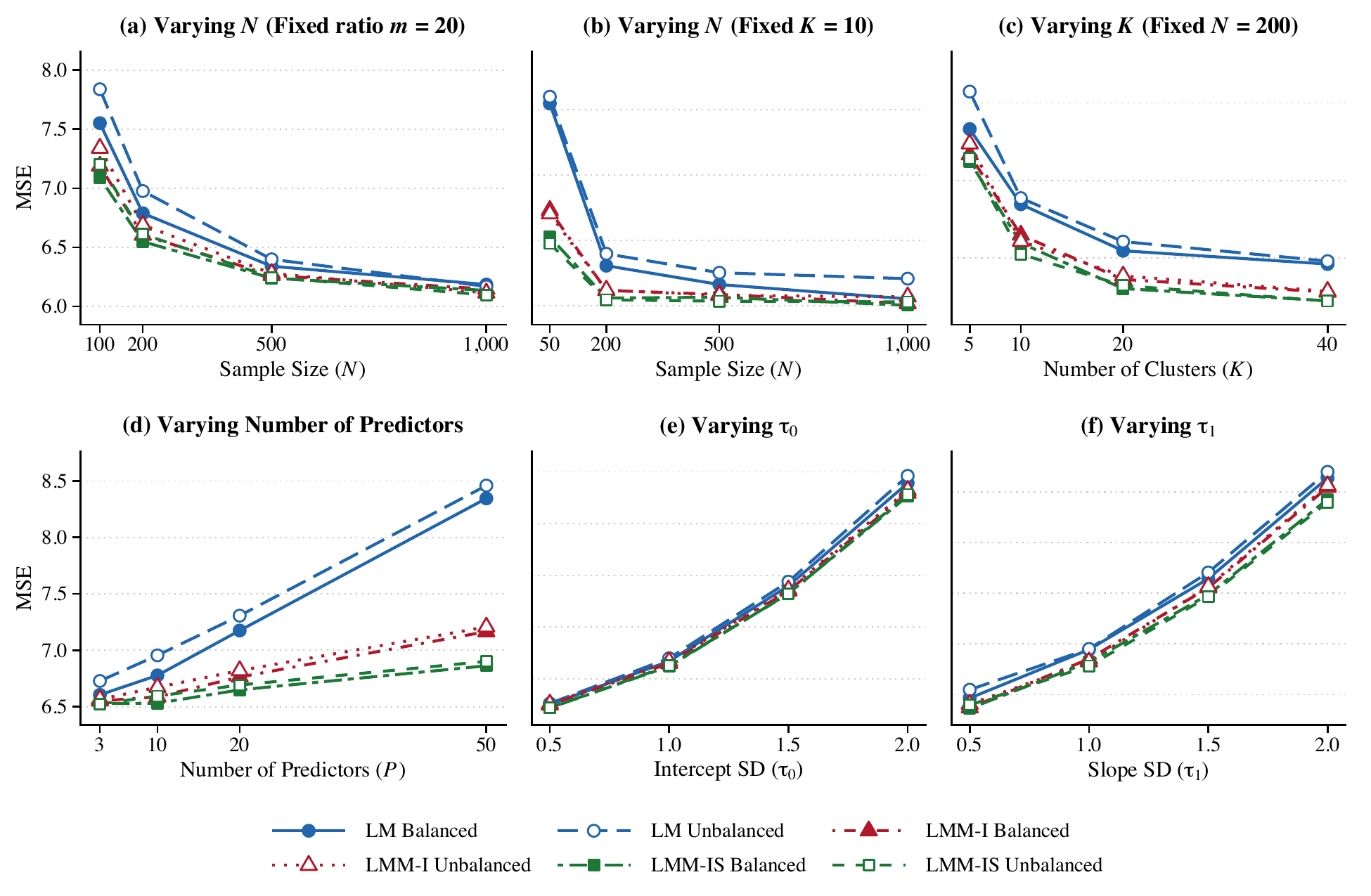}
\end{center}
\refstepcounter{figure}\label{fig:web-ris}\noindent\textbf{Web Figure \thefigure.} MSE for LM, LMM-I, and LMM-IS across random-intercept-and-random-slope simulation scenarios. Points show Monte Carlo means, and error bars show 95\% Monte Carlo confidence intervals.

\par\noindent\textbf{Alt text:} Multi-panel comparison of LM, random-intercept LMM, and random-intercept-and-slope LMM prediction MSE across simulation settings. The correctly specified random-intercept-and-slope model generally has the lowest MSE, the random-intercept model is intermediate, and the LM has the highest MSE; the additional benefit of the random slope increases with slope heterogeneity.

\end{document}